\documentclass[]{article}
\usepackage{amssymb}
\begin{document}

\title{Continuum percolation of simple fluids: Energetic connectivity criteria.}
\author{ Luis A. Pugnaloni \\ Procter Department of Food Science, University of Leeds, \\ Leeds LS2 9JT, UK \and Ileana F. M\'{a}rquez and Fernando Vericat \\ Instituto de F\'\i sica de L\'\i quidos y Sistemas Biol\'{o}gicos\\
(IFLYSIB)-UNLP-CONICET cc. 565 - (1900) La Plata, Argentina}

\maketitle 

\begin{abstract}
\smallskip
During the last few years, a number of works in computer simulation have focused on the clustering and percolation properties of simple fluids based in an energetic connectivity criterion proposed long ago by T.L. Hill [J. Chem. Phys. {\bf 23}, 617 (1955)]. This connectivity criterion appears to be the most appropriate in the study of gas-liquid phase transition. So far, integral equation theories have relayed on a velocity-averaged version of this criterion. We show, by using molecular dynamics simulations, that this average strongly overestimates percolation densities in the Lennard-Jones fluid making unreliable any prediction based on it. Additionally, we use a recently developed integral equation theory [Phys. Rev. E {\bf 61}, R6067 (2000)] to show how this velocity-average can be overcome.
\end{abstract}

\section{Introduction}

Clustering and percolation in continuum systems are concepts of great interest in chemical physics. Phenomena such as nucleation \cite{Senger1}, hydrogen bonding \cite{Starr1}, conductor-insulator transitions \cite{Chen1}, sol-gel transitions \cite{Butler1} and bridging in granular materials \cite{Pugnaloni0} are frequently studied in terms of clusters and percolation \cite{Stauffer}.

The first application of statistical mechanics formalism to describe clustering in equilibrium classical systems was done by Hill \cite{Hill1}. In Hill's theory, the concept of cluster is directly related to the idea of bonded-pairs. A bonded-pair is a set of two particles that are linked by some direct mechanism. Then, a cluster is a set of particles such that any pair of particles in the set is connected through a path of bonded-pairs. We call this clusters ``chemical clusters'' to distinguish them from the non-pair-bonded clusters we have introduced in a previous work \cite{Pugnaloni1}. A system is in a percolated state if it contains a cluster that spans the system.

From Hill's theory, we see that, in order to decide when two particles are bonded, a connectivity criterion is necessary. This connectivity criterion has to be defined according to the phenomenon under study \cite{comment1}. In the search for clusters of atoms in a monatomic gas that marks the onset of a phase transition, Hill proposed a simple energetic criterion: two particles are bonded if their relative kinetic energy is less than minus their relative potential energy \cite{Hill1}. However, this criterion is difficult to implement from a theoretical point of view and simpler ones were preferred. The very first simplification was set by Hill himself. Instead of using his general criterion, for which the relative positions and velocities of the relevant pair of particles come into account, he averaged out all the possible velocities of the particles. Then, a velocity-averaged (VA) criterion can be stated as follow: two particles are bonded with probability $P(r)$ if their relative potential energy is negative. $P(r)$ is calculated as an average probability over all possible relative velocities of a pair of particles \cite{Hill1}.

Coniglio {\it et. al.} have obtained an Ornstein-Zernike-type relationship for the pair connectedness function $g^{\dagger}({\bf r}_{1},{\bf r}_{2})$. This function is proportional to the joint probability density of finding two particles belonging to the same cluster and at positions ${\bf r}_{1}$ and ${\bf r}_{2}$, respectively \cite{Coniglio1}. Then, by integrating $g^{\dagger}({\bf r}_{1},{\bf r}_{2})$, the mean cluster size $S$ and the percolation density $\rho_{p}$ (i.e. the value of $\rho$ for which $S(\rho)$ diverges) can be obtained.

It is worth mentioning that most of the theoretical studies on connectivity and percolation in continuum systems (see for example Ref. \cite{Chiew1,DeSimone1,Laria1}) were focused in the rather simple Stillinger's connectivity criterion \cite{Stillinger1}. It states that two particles are bonded if they are separated by a distance shorter than a given connectivity distance $d$. In this case, $d$ is an {\it ad hoc} parameter, which must be chosen on physical grounds. Although this criterion might be sensible in the study of certain insulator-conductor transitions, it is unrealistic regarding clustering in saturated vapours.

More recently, Hill's criterion has been reconsidered in molecular dynamics studies of small clusters \cite{Soto1,Soto2} and the critical percolation behaviour of Lennard-Jones fluids \cite{Campi1}. It has been suggested that the percolation line $-$ the line that separates the temperature-density phase space into percolated and non-percolated states $-$ might be experimentally observable \cite{Campi1,Coniglio3}. Moreover, cluster analysis based in Hill's criterion seems to be useful in locating gas$-$liquid coexistence curves \cite{Campi1}. Notice that molecular dynamics simulations are mandatory if Hill's criterion is used to identify clusters since Monte Carlo algorithms do not provide the velocities of the particles to decide whether the relative kinetic energy of the a given pair outweighs its relative potential energy.

Since Coniglio's theory deals only with the positions of the particles, Hill's criterion cannot be achieved. Instead, the VA criterion was used by Coniglio {\it et. al.} \cite{Coniglio2} to analytically calculate, for a potential made up of a hard core plus an attractive interaction, the percolation loci in a crude mean-field approximation. Although they obtained that the percolation line ended just below the critical point, we show that this is an artefact of an additional approximation they introduced, which is not very suitable for the low temperatures involved.

In this work we present the results of a molecular dynamics (MD) investigation of the percolation loci for a Lennard-Jones system by using the two previous energetic criteria, i.e. VA and Hill's energetic criteria. We add a maximum connectivity distance $d$ to the criteria to avoid unrealistic bonding. Two particles separated by a distance greater than this maximum distance are considered as non-bonded even if their relative velocity is by chance zero \cite{Soto2}. We show that the velocity average introduces an important overestimation on the predicted percolation density at all temperatures. Therefore, the approximation introduced by this velocity average is not sensible. As a consequence, a way to implement the full Hill's connectivity criterion in an integral equation approach becomes essential for the future of these theories. Here, we show how a recently developed integral equation theory for continuum percolation \cite{Pugnaloni2} can be used to solve this problem.

The rest of the paper is organised as follow. In Sec. 2 we describe the MD simulations and show the unsuitability of the velocity average in the energetic bonding criterion. In Sec. 3 we briefly discuss Coniglio's percolation theory and two examples of its application to the VA criterion. In Sec. 4 a generalized integral equation theory is used to illustrate how the full Hill's energetic criterion can be fitted into this formalism. We conclude with some consideration about the possible future developments of this approach.

\section{Molecular dynamics}

Molecular dynamics simulation of $500$ particles in a cubic box with
periodic boundary conditions in the $NVT$ ensemble was performed by a
leap-frog algorithm with velocity correction \cite{Allen1}. Particles
interact via the Lennard-Jones potential 
\begin{equation}
v(r)=4\varepsilon \left[ \left( \frac{\sigma }{r}\right) ^{12}-\left( \frac{%
\sigma }{r}\right) ^{6}\right].  \label{e1}
\end{equation}

Time step was chosen as $\Delta t^{*}=\Delta t\sigma ^{-1}\sqrt{%
k_{B}T/(\varepsilon m)}=0.01$. Quantities are averaged over $10^{3}$
configurations picked up each 50 $\Delta t$ after stabilization. A cut off
distance equal to $2.7\sigma $ was used in the pair potential.

We identify the clusters by using both, the Hill's criterion for the bonded
probability defined as \cite{Hill1}: 
\begin{equation}
P({\bf r}_{i,j},{\bf p}_{i,j})=\left\{ 
\begin{array}{ll}
1 & {\bf p}_{i,j}^{2}/4m<-v({\bf r}_{i},{\bf r}_{j})\:and\:r_{i,j}\leq d
\\ 
0 & {\bf p}_{i,j}^{2}/4m\geqslant -v({\bf r}_{i},{\bf r}_{j})\:or\:%
r_{i,j}>d
\end{array}
\right.  \label{e2}
\end{equation}
(${\bf r}_{i,j}$ and ${\bf p}_{i,j}$ are the relative position and momentum
of particles $i$ and $j)$, and the VA criterion, which
correspond to \cite{Hill1,Coniglio2}: 
\begin{equation}
P_{av}(r_{i,j})=\left\{ 
\begin{array}{cc}
0 & v(r_{i,j})>0\:or\:r_{i,j}>d \\ 
\gamma [3/2,-v(r_{i,j})/k_{B}T]/\Gamma [3/2] & v(r_{i,j})\leq 0\:and\:%
r_{i,j}\leq d
\end{array}
\right.  \label{e3}
\end{equation}
($\Gamma [a]$ is the gamma function and $\gamma [a,x]$ is the incomplete gamma function).

In the last case, for each pair of particles that satisfy $v(r_{i,j})\leq 0$ and $r_{i,j}\leq d$, we generate a random number $z$, between $0$ and $1$.
If $z<\gamma [3/2,-v(r_{i,j})/k_{B}T]/\Gamma [3/2]$ we consider that the particles form a bonded-pair, otherwise we do not. Note that this criterion can also be used in Monte Carlo simulations because it does not need the knowledge of the particle velocities. We also identify clusters according to Stillinger's criterion for the sake of comparison. To identify the clusters from the list of bonded-pairs we use the Stoddard algorithm \cite{Allen1,Stoddard1}.

A system is said to be in a percolated state if a cluster that spans through the replicas is present 50 percent of the time \cite{Seaton1}. Then, a percolation transition curve, which separates the percolated and the non-percolated states of the system, can be drawn above the coexistence curve in the $%
T-\rho $ phase diagram.

In Fig. 1 percolation loci for Hill's, VA, and Stillinger's connectivity criterion are presented for $d=2\sigma $. The coexistence curve obtained by Panagiotopoulos \cite{Panagio1} is also included for comparison. Theoretical predictions from connectedness integral equation theory are shown for VA and Stillinger's criteria only (see Sec. 3). The two distinctive differences of the results for the two energetic criteria, as compared with the Stillinger's criterion, are: i) a strong dependence on temperature of the percolation curves and ii) a shift towards higher densities.

As it can be seen in Fig. 1, the VA criterion yields higher percolation densities than Hill's criterion. As it is expected, the approximation due to the average is less important as temperature drops: for low temperatures, the kinetic energy is small compared to the potential energy and an average over velocities is a good approximation. Nevertheless, the overestimation of the percolation density is never small. Therefore, it is expected that any theory based on the VA criterion will fail in describing the proper percolation (and in general clustering) behaviour of simple fluids. Any success should be attributed to cancellation of errors rather than accuracy of the theory. Indeed, we shall see in the next section that a Percus-Yevick-like approximation in the connectedness Ornstein$-$Zernike integral equation yields, for the VA criterion, an apparently good percolation line as compared with the full Hill's criterion.

It is worth mentioning that the results for the Hill's criterion are similar to the ones obtained by Campi {\it et. al.} \cite{Campi1} in a more extensive MD simulation. However, they obtained a faster increase in the percolation density with increasing temperature. The differences between our results and those from Campi {\it et. al.} could be related to the small size of our system and most importantly to the criterion for deciding whether the system is percolated or not. In Ref. \cite{Campi1} the authors considered that the system is on the percolation line if the second moment of the cluster size distribution $n'(s)$ that excludes the largest cluster reaches its maximum. Nevertheless, we are interested in the suitability of the VA bonding criterion and we expect that any criterion to calculate the percolation line might be affected to a similar extent by this simplification.

\section{Continuum clustering theory for non-velocity-dependent bonding criteria}

\subsection{General Theory}

In order to study clustering in a system composed of $N$ classical particles interacting via a pair potential $v({\bf r}_{1},{\bf r}_{2})$, Hill separated the Boltzmann factor $e({\bf r}_{1},{\bf r}_{2})=\exp[-\beta v({\bf r}_{1},{\bf r}_{2})]$, into bonded $(\dagger)$ and unbounded $(*)$ terms \cite{Hill1}: $e({\bf r}_{1},{\bf r}_{2})= e^\dagger({\bf r}_{1},{\bf r}_{2})+ e^*({\bf r}_{1},{\bf r}_{2})$. As usual $\beta=1/k_BT$, being $k_B$ the Boltzmann constant. Since $e^\dagger({\bf r}_{1},{\bf r}_{2})$ represents the basic probability density that two particles at position ${\bf r}_{1}$ and ${\bf r}_{2}$ are bonded, this separation yields a diagrammatic expansion for the partition function in terms of ``chemical" clusters. For convenience we express Hill's separation as follow:

\begin{equation}
e^\dagger({\bf r}_{1},{\bf r}_{2})=P({\bf r}_{1},{\bf r}_{2}) \exp[-\beta v({\bf r}_{1},{\bf r}_{2})] \label{e3.1}
\end{equation}

\begin{equation}
e^*({\bf r}_{1},{\bf r}_{2})=[1-P({\bf r}_{1},{\bf r}_{2})] \exp[-\beta v({\bf r}_{1},{\bf r}_{2})] \label{e3.2}
\end{equation}
where $P({\bf r}_{1},{\bf r}_{2})$ is given by Eq. (\ref{e3}) in the case of the VA energetic criterion.

Fugacity and density expansions have been found, within Hill's formalism, by Coniglio and co-workers \cite{Coniglio1} for the pair connectedness function $g^{\dagger}({\bf r}_{1},{\bf r}_{2})$. This function is proportional to the joint probability density of finding two particles belonging to the same cluster and at positions ${\bf r}_{1}$ and ${\bf r}_{2}$, respectively. Moreover, by collecting nodal and non-nodal diagrams in these expansions an Ornstein-Zernike-type relationship is obtained

\begin{equation}
g^{\dagger }({\bf r}_{1},{\bf r}_{2})=c^{\dagger }({\bf r}_{1},{\bf r}%
_{2})+\rho \int c^{\dagger }({\bf r}_{1},{\bf r}_{3})g^{\dagger }({\bf r}%
_{3},{\bf r}_{2})d{\bf r}_{3}.  \label{e6}
\end{equation}

The function $ c^\dagger({\bf r}_{1},{\bf r}_{2})$ is the direct pair connectedness function. By posing a closure relation between this function and $g^{\dagger}({\bf r}_{1},{\bf r}_{2})$, Eq. (\ref{e6}) can be solved. Then, the mean cluster size can be obtained as

\begin{equation}
S(\rho)=1+ \frac{1}{(N-1)} \int g^{\dagger}({\bf r}_{1},{\bf r}_{2}) d{\bf r}_{1} d{\bf r}_{2} \label{e3.3}
\end{equation}
were the integration is carried out over the whole system volume.

Percolation of an ensemble of particles is concerned with the existence of clusters that become macroscopic in size. Therefore, the percolation transition occurs at a critical percolation density $\rho _{c}$ which mathematically verifies

\begin{equation}
lim_{\!\!\!\!\!\!\!\!\!\!\!\!_{\rho \rightarrow \rho_{c}}} S(\rho)=\infty. \label{e3.4}
\end{equation}

In the rest of this section we present results for two systems were the VA criterion is applied.

\subsection{A simple example}

A theoretical calculation of the percolation loci for the VA criterion [see Eq. (\ref{e3})] was done by Coniglio {\it et. al.} \cite{Coniglio2} for a system in which the pair potential has the following form:

\begin{equation}
v({\bf r}_{1},{\bf r}_{2})=v(r_{1,2})=\left\{ 
\begin{array}{ll}
\infty & r_{1,2}\leq r_{0} \\ 
-u_{0}\left( \frac{r_{0}}{r_{1,2}}\right) ^{6} & r_{1,2}>r_{0}
\end{array}
\right. .  \label{e4}
\end{equation}
However, they did not use a maximum connectivity distance $d$ to avoid unphysical connections. Moreover, they used an expansion for the incomplete gamma function up to first order. We have recalculated numerically the percolation curve using $d=2\sigma $ and the complete expansion of the gamma function. The corresponding expression for the percolation density $\rho _{p}$ is \cite{Coniglio2}: 
\begin{equation}
\rho _{p}=\frac{1}{\overline{c}^{\dagger }(k=0)},  \label{e5}
\end{equation}
where $\overline{c}^{\dagger }(k)$ is the Fourier transform of the direct pair connectedness function.  We have used the closure relation for Eq. (\ref{e6}) proposed by Coniglio {\it et. al.,} i.e.

\begin{equation}
c^{\dagger }({\bf r}_{1},{\bf r}_{2})=f^{\dagger }(r_{1,2})  \label{e7}
\end{equation}
where $f^{\dagger }(r_{i,j})=e^{\dagger }(r_{i,j})=\exp [-\beta v(r_{i,j})]P_{av}(r_{i,j})$ is the basic probability for finding two particles separated by a distance $r_{i,j}$ and  bonded.  The function $f^{\dagger }(r_{i,j})$ is called the bound Mayer function.

In Fig. 2 we reproduce the coexistence curve and the percolation line, for the VA criterion, from Ref. \cite{Coniglio2}. The corrected results, using the complete expansion for the incomplete gamma function and the maximum connectivity distance $d$, are also included. Additionally, we show the percolation line for the full Hill's criterion (see Sec. 4). As we can see, using the full expansion of the incomplete gamma function is important since results are very much affected. The importance of the maximum connectivity distance $d$ is negligible. We calculated the percolation line for increasing values of $d$ $-$ up to infinity $-$ without obtaining significant changes.

Based on the earlier results, it was suggested \cite{Coniglio3} that this theory is rather satisfactory considering its simplicity because the predicted percolation line ended close to the critical point as MD simulations show \cite{Campi1} for the full Hill connectivity criterion. However, this might be due to a cancellation of errors. On the one hand, the actual percolation line for the VA criterion should be shifted to even higher densities as it can be seen in Fig. 1 for the Lennard-Jones system. On the other hand, the approximation for the incomplete gamma function seems to affect the result largely.

\subsection{Numerical results for the Lenard-Jones system}

In order to solve Eq. (\ref{e6}) for the Lennard-Jones potential we have implemented Labik's numerical algorithm \cite{Labik1}. We have already seen that the simple closure relation given by Eq. (\ref{e7}) is very crude. Hence, we use the more reliable closure available, {\it i.e.} the Percus-Yevick-like relation

\begin{equation}
g^{\dagger}(r_{1,2})=[f^*(r_{1,2})+1][ g^{\dagger}(r_{1,2})- c^{\dagger}(r_{1,2})]+\exp[\beta v(r_{1,2})]g(r_{1,2})f^{\dagger}(r_{1,2}). \label{e3.2.1}
\end{equation}
In Eq. (\ref{e3.2.1}), $f^*(r_{1,2})=e^*(r_{1,2})-1=\exp [-\beta v(r_{i,j})][1-P_{av}(r_{i,j})]-1$ is the unbound Mayer function. We recall that $P_{av}(r_{i,j})$ is given by Eq. (\ref{e3}) for the VA energetic criterion.

By using Labik's algorithm we solve the coupled system of equations (\ref{e6}) and (\ref{e3.2.1}). Then, mean cluster size can be obtained for given $T$ and $\rho$. For a given temperature, percolation density is determined by fitting the power law $S^{-1}\propto |\rho -\rho _{c}|^{\gamma }$ for $\rho \lesssim \rho _{c}$ \cite{Seaton1}. Although critical exponents obtained by fitting $S$ as a function of $\rho$ are difficult to obtain accurately in these cases, percolation densities are very reliable.

In Eq. (\ref{e3.2.1}) the ``thermal" pair correlation function $g(r_{1,2})$ is involved. We have obtained $g(r_{1,2})$ by numerically solving the ``thermal" Ornstein-Zernike equation closed with the ``thermal" Percus-Yevick approximation \cite{Hansen1}.

In Fig. 1 we show the results for the VA criterion and the Stillinger's criterion for $d=2\sigma$. As we can see, the results from Percus-Yevick theory agree rather well with MD simulations for Stillinger's criterion. However, for the VA criterion, the theory is unsatisfactory. Nevertheless, we see a good advance with respect to the very low percolation densities predicted in the previous subsection, for a similar system, by the closure relation (\ref{e7}). We believe that the main cause of the failure for the VA criterion is related to the ``input" function $g(r_{1,2})$. The value of this function is less well predicted by the ``thermal" Percus-Yevick approximation at these ``high" densities and temperatures. Notice that these types of theories were developed to describe the liquid state, particularly near the triple point. At low densities, as the involved for the Stillinger's criterion, these theories are valid even for temperatures above the critical point.

Although the previous theory seems to be inappropriate to describe the percolation transition for the VA criterion, it seems to agree better with the MD result for the full Hill's criterion. This is, of course, due to a cancellation of errors between the velocity average and the Percus-Yevick approximation. Since Hill's criterion produces a percolation line at intermediate densities, we should be able to use the Percus-Yevick approximation successfully to study this case. However, getting rid of the velocity average is the main task to accomplish. In the following section we show how this can be achieved.

\section{Continuum clustering theory for velocity-dependent bonding criteria}

\subsection{General theory}

In this section we summarize the main results of a theory we have presented elsewhere \cite{Pugnaloni2} to describe the clustering and percolation for bonding criteria that involve the velocity of the particles as well as their positions.

For a system of $N$ classical particles interacting via a pair potential $v({\bf r}_{i},{\bf r}_{j})$ we define a density correlation function $\rho ({\bf r}_{1},{\bf r}_{2},{\bf p}_{1},{\bf p}_{2})$ which is $N(N-1)$ times
the probability density of finding two particles at phase space points $({\bf r}_{1}$, ${\bf p}_{1})$ and $({\bf r}_{2}$, ${\bf p}_{2})$ respectively:

\begin{eqnarray}
\rho ({\bf r}_{1},{\bf r}_{2},{\bf p}_{1},{\bf p}_{2}) &=&\frac{N(N-1)}{h^{3N}N!Q_{N}(V,T)}  \label{n8} \\
&&\times \int \prod_{i=1}^{N}\exp [-\beta \frac{{\bf p}_{i}^{2}}{2m}]\prod_{i=1}^{N}\prod_{j>i}^{N}\exp [-\beta v({\bf r}_{i},{\bf r}_{j})]d{\bf r}^{N-2}d{\bf p}^{N-2}.  \nonumber
\end{eqnarray}
Here $h$ is the Planck constant and $Q_{N}(V,T)$ the canonical partition function of the system. Then, in the same spirit of Hill and Coniglio {\it et. al.} \cite{Hill1,Coniglio1}, we separate $\exp [-\beta v({\bf r}_{i},{\bf r}_{j})] $ into connecting and blocking parts

\begin{equation}
\exp [-\beta v({\bf r}_{i},{\bf r}_{j})]=f^{\dagger }({\bf r}_{i},{\bf r}_{j},{\bf p}_{i},{\bf p}_{j})+f^{*}({\bf r}_{i},{\bf r}_{j},{\bf p}_{i},{\bf p}_{j})+1  \label{n9}
\end{equation}
Here $f^{\dagger }({\bf r}_{i},{\bf r}_{j},{\bf p}_{i},{\bf p}_{j})$ represent the basic probability density that two particles at configuration $({\bf r}_{i},{\bf r}_{j},{\bf p}_{i},{\bf p}_{j})$, are bonded. The shorthand notation $f^{\gamma }({\bf r}_{i},{\bf r}_{j},{\bf p}_{i},{\bf p}_{j})\equiv f_{i,j}^{\gamma }$ ($\gamma =\dagger ,*$) will be sometimes used.

Substitution of Eq. (\ref{n9}) in Eq. (\ref{n8}) yields
 
\begin{eqnarray}
\rho ({\bf r}_{1},{\bf r}_{2},{\bf p}_{1},{\bf p}_{2}) &=&\frac{N(N-1)}{h^{3N}N!Q_{N}(V,T)}\exp [-\beta v({\bf r}_{1},{\bf r}_{2})]  \label{n10} \\
&&\times \int \prod_{i=1}^{N}\exp [-\beta \frac{{\bf p}_{i}^{2}}{2m}]\sum
\{\prod f_{i,j}^{\dagger }f_{k,l}^{*}\}dr^{N-2}dp^{N-2},  \nonumber
\end{eqnarray}
where the sum is carried out over all possible arrangements of products of functions $f_{i,j}^{\dagger}$ and $f_{k,l}^{*}$.

It should be noted that the functions $f_{i,j}^{\dagger }$ and $f_{i,j}^{*}$ can depend on momenta as well as on the positions of the two particles, but the sum of $f_{i,j}^{\dagger }$ and $f_{i,j}^{*}$ must be momenta independent in order that Eq. (\ref{n9}) be satisfied. Except by this last condition, the functions $f_{i,j}^{\dagger }$ and $f_{i,j}^{*}$ are otherwise arbitrary for thermodynamic purposes. Obviously, we choose them in such a way that the Hill's definition of bonded particles is achieved:

\begin{equation}
f_{i,j}^{\dagger }=\exp [-\beta v(r_{i,j})]P({\bf r}_{i,j},{\bf p}_{i,j})
\label{n11}
\end{equation}

\begin{equation}
f_{i,j}^{*}=\exp [-\beta v(r_{i,j})][1-P({\bf r}_{i,j},{\bf p}_{i,j})]-1
\label{n12}
\end{equation}
where $P({\bf r}_{i,j},{\bf p}_{i,j})$ is given in Eq. (\ref{e2}).

Each term in the integrand of Eq. (\ref{n10}) can be represented as a diagram
consisting of two white $e_{1}$ and $e_{2}$-points, $N-2$ black $e_{i}$%
-points and some $f_{i,j}^{\dagger }$ and $f_{i,j}^{*}$ connections except
between the white points. Here we take $e_{i}\equiv \exp [-\beta \frac{{\bf p%
}_{i}^{2}}{2m}]$. White points are not integrated over whereas black points
are integrated over their positions and momenta. All the machinery normally
used to handle standard diagrams in classical liquid theory \cite{Hansen1}
can be now extended to treat these new kind of diagrams. By following
Coniglio's recipe to separate connecting and blocking parts in correlation
functions i.e. $g({\bf r}_{1},{\bf r}_{2})=g^{\dagger }({\bf r}_{1},{\bf r}_{2},{\bf p}_{1},{\bf p}_{2})+g^{*}({\bf r}_{1},{\bf r}_{2},{\bf p}_{1},{\bf p}_{2})$, we obtain an Ornstein-Zernike like integral equation for $g^{\dagger }({\bf r}_{1},{\bf r}_{2},{\bf p}_{1},{\bf p}_{2})$\begin{eqnarray}
g^{\dagger }({\bf r}_{1},{\bf r}_{2},{\bf p}_{1},{\bf p}_{2}) &=&c^{\dagger
}({\bf r}_{1},{\bf r}_{2},{\bf p}_{1},{\bf p}_{2})  \label{n13} \\
&&+\int \rho ({\bf r}_{3},{\bf p}_{3})c^{\dagger }({\bf r}_{1},{\bf r}_{3},{\bf p}_{1},{\bf p}_{3})g^{\dagger }({\bf r}_{3},{\bf r}_{2},{\bf p}_{3},{\bf p}_{2})d{\bf r}_{3}d{\bf p}_{3}.  \nonumber
\end{eqnarray}

Here $\rho ({\bf r}_{1},{\bf p}_{1})\rho ({\bf r}_{2},{\bf p}_{2})g^{\dagger
}({\bf r}_{1},{\bf r}_{2},{\bf p}_{1},{\bf p}_{2})$ is $N(N-1)$ times the
joint probability density of finding two particles at positions ${\bf r}_{1}$
and ${\bf r}_{2}$ with momenta ${\bf p}_{1}$ and ${\bf p}_{2}$,
respectively, and belonging to the same cluster, where the bonding criterion
is given by Eqs. (\ref{n11}) and (\ref{n12}); being 
\begin{equation}
\rho ({\bf r}_{1},{\bf p}_{1})=\frac{1}{N-1}\int \rho ({\bf r}_{1},{\bf r}_{2},{\bf p}_{1},{\bf p}_{2})d{\bf r}_{2}d{\bf p}_{2}.  \label{n14}
\end{equation}
The function $c^{\dagger }({\bf r}_{1},{\bf r}_{2},{\bf p}_{1},{\bf p}_{2})$
denotes the sum of all the non-nodal diagrams in the diagrammatic expansion
of $g^{\dagger }({\bf r}_{1},{\bf r}_{2},{\bf p}_{1},{\bf p}_{2}).$ We
remember that a nodal diagram contains at least a black point\thinspace
through which all paths between the two white points pass.

For homogeneous systems we have

\begin{eqnarray}
g^{\dagger }({\bf r}_{1},{\bf r}_{2},{\bf p}_{1},{\bf p}_{2}) &=&c^{\dagger
}({\bf r}_{1},{\bf r}_{2},{\bf p}_{1},{\bf p}_{2})+\frac{\rho }{(2\pi
mk_{B}T)^{3/2}}  \label{n15} \\
&&\times \int \exp [-\beta \frac{{\bf p}_{3}^{2}}{2m}]c^{\dagger }({\bf r}_{1},{\bf r}_{3},{\bf p}_{1},{\bf p}_{3})g^{\dagger }({\bf r}_{3},{\bf r}_{2},{\bf p}_{3},{\bf p}_{2})d{\bf r}_{3}d{\bf p}_{3}.  \nonumber
\end{eqnarray}

To get an integral equation from Eq. (\ref{n13}) is necessary a closure
relation between $g^{\dagger }({\bf r}_{1},{\bf r}_{2},{\bf p}_{1},{\bf p}_{2})$ and $c^{\dagger }({\bf r}_{1},{\bf r}_{2},{\bf p}_{1},{\bf p}_{2})$.
As an example, we can take the Percus-Yevick approximation $g({\bf r}_{1},{\bf r}_{2})\exp
[\beta v({\bf r}_{1},{\bf r}_{2})]=1+N({\bf r}_{1},{\bf r}_{2}),$ where the
function $N({\bf r}_{1},{\bf r}_{2})$ is the sum of the nodal diagrams in
the expansion of $g({\bf r}_{1},{\bf r}_{2})$. Separation into connecting
and blocking parts i.e. $g({\bf r}_{1},{\bf r}_{2})=g^{\dagger }({\bf r}_{1},{\bf r}_{2},{\bf p}_{1},{\bf p}_{2})+g^{*}({\bf r}_{1},{\bf r}_{2},{\bf p}_{1},{\bf p}_{2})$ and $N({\bf r}_{1},{\bf r}_{2})=N^{\dagger }({\bf r}_{1},{\bf r}_{2},{\bf p}_{1},{\bf p}_{2})+N^{*}({\bf r}_{1},{\bf r}_{2},{\bf p}_{1},{\bf p}_{2})$, yields 
\begin{eqnarray}
g^{\dagger }({\bf r}_{1},{\bf r}_{2},{\bf p}_{1},{\bf p}_{2}) &=&[f^{*}({\bf r}_{1},{\bf r}_{2},{\bf p}_{1},{\bf p}_{2})+1][g^{\dagger }({\bf r}_{1},{\bf r}_{2},{\bf p}_{1},{\bf p}_{2})-c^{\dagger }({\bf r}_{1},{\bf r}_{2},{\bf p}_{1},{\bf p}_{2})] \nonumber \\
&&+\exp [\beta v({\bf r}_{1},{\bf r}_{2})]g({\bf r}_{1},{\bf r}_{2})f^{\dagger }({\bf r}_{1},{\bf r}_{2},{\bf p}_{1},{\bf p}_{2}). 
\label{n16} 
\end{eqnarray}
Eq. (\ref{n13}) closed by Eq. (\ref{n16}) gives an integral equation for $%
g^{\dagger }({\bf r}_{1},{\bf r}_{2},{\bf p}_{1},{\bf p}_{2})$.

From the function $g^{\dagger }({\bf r}_{1},{\bf r}_{2},{\bf p}_{1},{\bf p}%
_{2})$ we define the pair connectedness function

\begin{equation}
g^{\dagger}({\bf r}_{1},{\bf r}_{2})=\int \rho ({\bf r}_{1},{\bf p}_{1})\rho ({\bf r}_{2},{\bf p}_{2})g^{\dagger }({\bf r}_{1},{\bf r}_{2},{\bf p}_{1},{\bf p}_{2})d{\bf p}_{1}d{\bf p}_{2}.  \label{n17}
\end{equation}
This function is the joint probability density of finding two particles that belong to
the same cluster, within the Hill's criterion, at positions
${\bf r}_{1}$ and ${\bf r}_{2}$, respectively.

Then, the mean cluster size $S$ is given by

\begin{equation}
S=1+\frac{1}{(N-1)}\int g^{\dagger}({\bf r}_{1},{\bf r}_{2})d{\bf r}_{1}d{\bf r}_{2}. 
\label{n18}
\end{equation}

\subsection{A simple example}

By using the previous extension of Coniglio's work, we have calculated the percolation loci for the full Hill's criterion. We have used the pair potential of Eq. (\ref{e4}). In this case we solve Eq. (\ref{n15}) with a closure relation equivalent to Eq. (\ref{e7}), i.e.

\begin{equation}
c^{\dagger }({\bf r}_{1},{\bf r}_{2},{\bf p}_{1},{\bf p}_{2})=f^{\dagger}
({\bf r}_{1},{\bf r}_{2},{\bf p}_{1},{\bf p}_{2}).  \label{e9}
\end{equation}
where $f^{\dagger}
({\bf r}_{1},{\bf r}_{2},{\bf p}_{1},{\bf p}_{2})$ is given by Eq. (\ref{n11}). Then

\begin{equation}
c^{\dagger }({\bf r}_{1},{\bf r}_{2},{\bf p}_{1},{\bf p}_{2})=\left\{ 
\begin{array}{ll}
0 & r_{1,2}\leq r_{0}\:or\:(p_{1,2})^{2}/4m\geq -v(r_{1,2}) \\ 
\exp \left[ \beta u_{0}\left( \frac{r_{0}}{r_{1,2}}\right) ^{6}\right] & 
r_{1,2}>r_{0}\:and\:(p_{1,2})^{2}/4m<-v(r_{1,2})
\end{array}
\right. .  \label{e10}
\end{equation}

In order to obtain the mean cluster size we solve the system (\ref{n15}) and (\ref{e10}) as follow. Firstly, taken into account the homogeneity of the system, we perform a Fourier transform on Eq. (\ref{n15}) with respect to ${\bf r}_{1,2}$
 
\begin{eqnarray}
g^{\dagger }({\bf k}_{1,2},{\bf p}_{1},{\bf p}_{2}) &=&c^{\dagger }({\bf k}_{1,2},{\bf p}_{1},{\bf p}_{2})+\frac{\rho }{(2\pi mk_{B}T)^{3/2}}
 \label{e11} \\
&&\times \int \exp [-\beta \frac{{\bf p}_{3}^{2}}{2m}]c^{\dagger }({\bf k}_{1,2},{\bf p}_{1},{\bf p}_{3})g^{\dagger }({\bf k}_{1,2},{\bf p}_{3},{\bf p}_{2})d{\bf p}_{3}  \nonumber
\end{eqnarray}
where
 
\begin{equation}
g^{\dagger }({\bf k}_{1,2},{\bf p}_{1},{\bf p}_{2})=\int g^{\dagger }({\bf r}_{1,2},{\bf p}_{1},{\bf p}_{2})\exp [-i{\bf r}_{1,2}{\bf k}_{1,2}]d{\bf r}_{1,2}  \label{e12}
\end{equation}
and an analogous expression for $c^{\dagger }({\bf k}_{1,2},{\bf p}_{1},{\bf p}_{2})$.

By evaluating the previous equation in ${\bf k}_{1,2}=0$, multiplying by $%
\exp [-\beta \frac{{\bf p}_{2}^{2}}{2m}]$ and integrating with respect to $%
{\bf p}_{2}$ we obtain 
\begin{eqnarray}
g_{0}^{\dagger }({\bf p}_{1}) &=&c_{0}^{\dagger }({\bf p}_{1})+\frac{\rho }{(2\pi mk_{B}T)^{3/2}}\times  \label{e13} \\
&&\int \exp [-\beta \frac{{\bf p}_{3}^{2}}{2m}]c_{0}^{\dagger }({\bf p}_{1},{\bf p}_{3})g_{0}^{\dagger }({\bf p}_{3})d{\bf p}_{3}  \nonumber
\end{eqnarray}
with 
\begin{equation}
g_{0}^{\dagger }({\bf p}_{1},{\bf p}_{2})\equiv g^{\dagger }({\bf k}_{1,2}=0,{\bf p}_{1},{\bf p}_{2}),  \label{e14}
\end{equation}
\begin{equation}
g_{0}^{\dagger }({\bf p}_{1})=\int \exp [-\beta \frac{{\bf p}_{2}^{2}}{2m}]g_{0}^{\dagger }({\bf p}_{1},{\bf p}_{2})d{\bf p}_{2},  \label{e15}
\end{equation}
and analogous expressions for $c^{\dagger }({\bf k}_{1,2},{\bf p}_{1},{\bf p}_{2})$.

Because $c_{1,2}^{\dagger }$ in Eq. (\ref{e10}) depends just on the modulus
of ${\bf r}_{1}-{\bf r}_{2}$ and ${\bf p}_{1}-{\bf p}_{2}$, we can perform a
Fourier transform with respect to ${\bf p}_{1}$ yielding 
\begin{equation}
g_{0}^{\dagger }({\bf \omega })=\left( \frac{2\pi m}{\beta }\right)
^{3/2}\exp [-\frac{3m\omega ^{2}}{2\beta }]c_{0}^{\dagger }({\bf \omega })+
\frac{\rho }{(2\pi mk_{B}T)^{3/2}}c_{0}^{\dagger }({\bf \omega })\hat{g}_{0}^{\dagger }({\bf \omega })  \label{e16}
\end{equation}
where we have defined 
\begin{equation}
g_{0}^{\dagger }({\bf \omega })=\int g_{0}^{\dagger }({\bf p}_{1})\exp [-i{\bf \omega p}_{1}]d{\bf p}_{1}  \label{e17}
\end{equation}
\begin{equation}
\hat{g}_{0}^{\dagger }({\bf \omega })=\int g_{0}^{\dagger }({\bf p}_{3})\exp
[-\beta \frac{{\bf p}_{3}^{2}}{2m}]\exp [-i{\bf \omega p}_{3}]d{\bf p}_{3}
\label{e18}
\end{equation}
\begin{equation}
c_{0}^{\dagger }({\bf \omega })=\int c_{0}^{\dagger }(\left| {\bf p}_{1}-{\bf p}_{3}\right| )\exp [-i{\bf \omega p}_{1,3}]d{\bf p}_{1,3}.  \label{e19}
\end{equation}

From Eq. (\ref{n18}), the mean cluster size can be expressed as 
\begin{equation}
S=1+\frac{1}{(N-1)}\int \rho ({\bf r}_{1},{\bf p}_{1})\rho ({\bf r}_{2},{\bf p}_{2})g^{\dagger }({\bf r}_{1},{\bf r}_{2},{\bf p}_{1},{\bf p}_{2})d{\bf p}_{1}d{\bf p}_{2}d{\bf r}_{1}d{\bf r}_{2},  \label{e20}
\end{equation}
or, as a function of $\hat{g}_{0}^{\dagger }({\bf \omega }),$ 
\begin{equation}
S=1+\frac{\rho }{(2\pi mk_{B}T)^{3}}\hat{g}_{0}^{\dagger }({\bf \omega =0}).
\label{e21}
\end{equation}

Taking $g_{0}^{\dagger }({\bf \omega })=\left( \frac{2\pi m}{\beta }\right)
^{3/2}\exp [-\frac{3m\omega ^{2}}{2\beta }]c_{0}^{\dagger }({\bf \omega })$
as a first order approximation and iterating Eq. (\ref{e16}) up to
convergence we can calculate the mean cluster size for given $\rho $ and $T.$

As in the previous section, percolation density is determined by fitting the
power law $S^{-1}\propto |\rho -\rho _{c}|^{\gamma }$ for $\rho \lesssim
\rho _{c}$.

In Fig. 2, we present the percolation line obtained for Hill's criterion. As for the VA criterion, the simple closure relation implemented yields unrealistic percolation densities. However, it can be seen that the percolation line is shifted to the left, as compared with the one corresponding to the VA criterion. This line is also less temperature dependent $-$ it presents a sharper slope in the $T-\rho$ plane. These trends are consistent with the MD simulation for the Lennard-Jones fluid (see Fig. 1). The discrepancy between both criteria is negligible only at low temperature as expected.
 
Although a satisfactory theory for the continuum percolation of a simple fluid within the Hill's energetic connectivity criterion is still due, we believe the main step forward has been presented here. The next attempt to improve the theory should be the implementation of the Percus-Yevick connectedness closure relation (\ref{n16}). This can be done by solving numerically the coupled system (\ref{n15}) and (\ref{n16}). This is a non-straightforward task: tensors of rank 9 must be used to storage correlation functions and a double convolution has to be solved. Recently, F. Lado \cite{Lado1} has developed a method to calculate the pair correlations for a polarizable Lennard-Jones system which is based in an expansion of the relevant functions in orthogonal polynomials. The integral equation he has to solve is similar to Eq. (\ref{n13}) but with ${\bf p}_{1}$ and ${\bf p}_{2}$ representing the induced dipolar moments of the particles 1 and 2 respectively. A modification of this method might be used for the clustering problem of Lennard-Jones particles by using the Percus-Yevick like closure of Eq. (\ref{n16}).

\section{Concluding remarks}

In this work we have shown that the common introduction of an average over velocities in the connectivity energetic criteria for simple fluids strongly overestimates the percolation density. Moreover, we have shown how this average can be avoided by using a more general continuum percolation theory. Although this theory was initially developed to implement time-dependent connectivity criteria, it is specially suited to the Hill's energetic criteria.

Even though we have applied the new formalism to avoid the velocity-average, we used a very simple approximation to close the generalized Ornstein-Zernike connectedness equation. Nevertheless, the results are promising in that the trends obtained by MD simulation are reproduced. A more sophisticated closure relation should be implemented in order to obtain a more satisfactory theory.
 
\bigskip

\bigskip \bigskip \bigskip \bigskip

{\bf ACKNOWLEDGMENTS}

Support of this work by UNLP (Grant I055), CONICET (PIP96 4690) and ANPCYT
(PICT 03-04517) of Argentina is very much appreciated. I.F.M. acknowledge fellowships from ANPCYT and CONICET. F.V. is a member of CONICET.

\begin{center}
{\bf Figure Captions}
\end{center}

{\bf Figure 1:} Coexistence and percolation curves for the Lennard-Jones fluid: MD (symbols) and Percus-Yevik connectedness theory (lines). Densities and temperatures are in units of $\sigma^{-3}$ and $\varepsilon /k_{B}$, respectively. Solid diamonds (with trend line) correspond to the coexistence curve. Percolation loci for Stillinger's criterion (open squares and dotted line), Hill's criterion (open circles), and VA criterion (open triangles and solid line) are presented.
 
{\bf Figure 2:} Coexistence and percolation lines for a simple fluid. Full and dashed lines correspond, respectively, to the percolation loci for Hill's criterion [calculated from Eqs. (\ref{n15}) and (\ref{e10})] and VA criterion [calculated from Eqs. (\ref{e6}) and (\ref{e7})]. Dotted line for VA criterion is reproduced from Ref. \cite{Coniglio2} to be compared with the ``corrected" results (dashed line).
 The coexistence line is also reproduced from Ref. \cite{Coniglio2}. The particles of the fluid interact through the pair potential given by Eq. (\ref{e4}). Densities and temperatures are reduced with the corresponding critical values.


\begin{thebibliography}{60}
\bibitem{comment1}  For example, in the conductor-insulator transition of water in oil microemulsions two micelles are consider bonded if they can share charge carriers (see Ref. \cite{Chen1}). However, in the identification of bridges formed by grains within granular matter, two grains are bonded if they stabilize each other (see Ref. \cite{Pugnaloni0}).
\end{thebibliography}
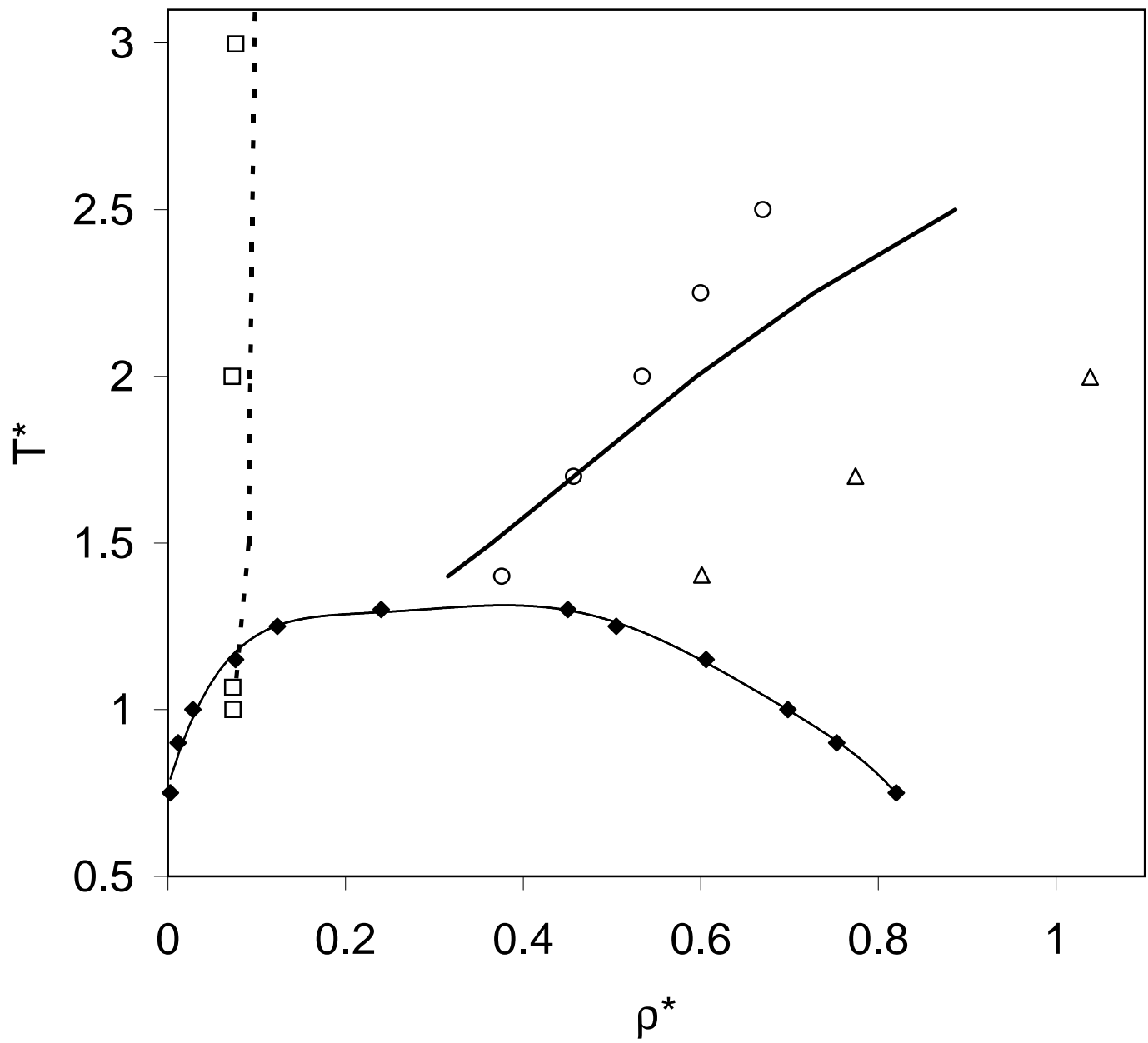
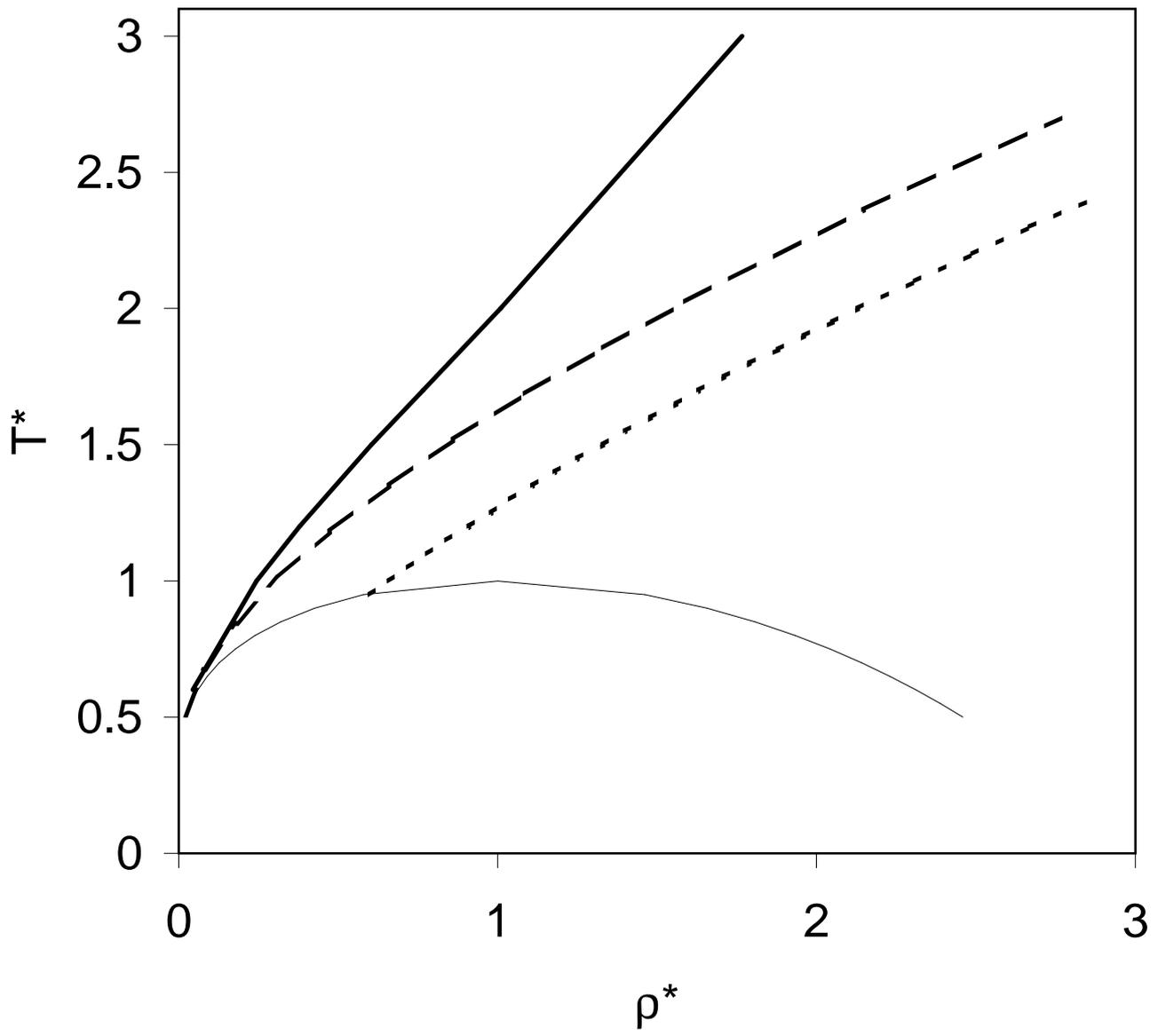
\end{document}